# Deep learning neural network for approaching Schrödinger problems with arbitrary two-dimensional confinement

A. Radu[1*], C. A. Duque[2]

[1]Department of Physics, Politehnica University of Bucharest, 313 Splaiul Independenței, Bucharest, RO-060042, Romania

[2]Grupo de Materia Condensada-UdeA, Instituto de Física, Facultad de Ciencias Exactas y Naturales, Universidad de Antioquia UdeA, Calle 70 No. 52-21, Medellín, Colombia

*Corresponding author: adrian.radu@physics.pub.ro

**Abstract**

This article presents an approach to the two-dimensional Schrödinger equation based on automatic learning methods with neural networks. It is intended to determine the ground state of a particle confined in any two-dimensional potential, starting from the knowledge of the solutions to a large number of arbitrary sample problems. A network architecture with two hidden layers is proposed to predict the wave function and energy of the ground state. Several accuracy indicators are proposed for validating the estimates provided by the neural network. The testing of the trained network is done by applying it to a large set of confinement potentials different from those used in the learning process. Some particular cases with symmetrical potentials are solved as concrete examples, and a good network prediction accuracy is found.

## 1. Introduction

The speed of computerized data processing and the ability to analyze large datasets have increased exponentially as a direct consequence of the rapid development of processors and computing techniques. This evolution proved important not only from a quantitative perspective but also led to a paradigm shift in terms of methods and algorithms for solving difficult problems. In recent years, more and more categories of concrete technical problems but also of fundamental interest problems have been addressed by new artificial intelligence methods, such as machine learning (ML) [1]. A concrete way in which this method can be implemented is with the help of neural networks (NNs), logical structures inspired by the functioning of the biological nervous system. With this approach, the problems of automatic classification or recognition of shapes that are almost unapproachable by classical algorithms can be solved. However, various more abstract problems from fundamental disciplines have proved to be approachable from this new perspective. Using NNs to solve problems that already admit more or less sophisticated conventional solutions can also be instructive and present the potential for further extensions to more complicated contexts. One such relatively recent challenge of interest in nanomaterials science and quantum chemistry is to solve the Schrödinger Equation (SE) in one or more dimensions using artificial intelligence methods. Some advantages of using ML methods compared to existing numerical methods for quantum physics are: obtaining much faster estimates of the energy of particles confined in nontrivial potentials, better approaching many-body problems, and predicting phase transitions and properties of quantum systems in inaccessible physical conditions.

ML predictive approaches, generally built upon statistical learning theory, represent a different paradigm from the classical methods for solving SE, although they can be based on their results in the training stage. Exact analytical solutions of SE for nanostructures can be obtained in very few simple cases and are therefore of little practical relevance [2]. Approximate quasi-analytical solutions can be obtained using variational techniques [3] or perturbative methods [4], which have been extensively studied in the last century [5]. However, they are limited in accuracy and impractical for physical systems with nontrivial geometries, dimensionalities, and interactions. Asymptotic iteration methods can be an alternative for solving 1-D Schrödinger-type problems [6,7]. Meshless methods and diagonalization techniques provide good results; however, they are more demanding in terms of computational effort [8,9]. In parallel with the accelerated development of computers, numerical methods based on spatial discretizations and finite-difference approximations of SE have been increasingly used [10-12]. Shooting methods use iterative solving of finite-difference equations with discrete energy values in a search interval and the selection of those that best meet the boundary conditions [13,14]. For all SE dimensionalities, an accurate and versatile mesh-based approach is the finite element method, which uses a weak formulation of the equation and involves large algebraic systems [15-20].

So far, there have been a few studies dealing with artificial intelligence methods for solving SE, the most important of which are mentioned below.

Lagaris *et al.* demonstrated how artificial NNs can be used to solve partial differential equations [21] and applied the concepts for solving SE in several cases with different dimensionalities [22]. Sugawara proposed a new approach for solving one-dimensional (1-D) SE by combining a genetic algorithm and an NN [23]. In a remarkable study, Mills *et al.* introduced a deep learning method for solving two-dimensional (2-D) SE by calculating the ground and first excited states of an electron in different types of confining potentials (CPs) [24]. Vargas-Hernández *et al.* presented an ML method based on Gaussian process regression to predict sharp transitions in a Hamiltonian phase diagram by extrapolating the properties of quantum systems [25]. Han *et al.* solved many-electron SE using a deep NN with wave function (WF) optimization through a Monte Carlo approach [26]. Using NNs, Mutuk addressed the eigenvalue problem of a 1-D anharmonic oscillator [27]. Manzhos reviewed recent ML techniques used to solve electronic and vibrational SEs, which are typically related to computational chemistry [28]. Hermann *et al.* proposed PauliNet, which is a deep NN representation of electronic WFs for molecules with up to 30 electrons, and proved that it can outperform variational quantum chemistry models [29]. Pfau *et al.* introduced a novel deep learning architecture, the Fermionic NN, to approach many-electron SE [30]. Li *et al.* used an NN model to solve SE by computing multiple excited states [31]. Grubišić *et al.* used a dense deep NN and a fully convolutional NN to approximate eigenmodes localized by a CP [32]. Yüksel *et al.* applied multilayer perceptron architectures to predict the ground-state binding energies of atomic nuclei [33]. In a study by da Silva Macedo *et al.* an NN was trained to predict the energy levels and energy-dependent masses as nonparabolic properties of semiconductor heterostructures [34]. The learning ability of a physics-informed proper orthogonal decomposition-Galerkin simulation methodology for QD structures was investigated by Veresko and Cheng [35]. In a recently published paper, we used two different neural architectures to approach 1-D SE in quantum wells (QWs) with arbitrary CPs [36]. The results were compared and discussed using accuracy indicators and represent the starting point of the 2-D generalization addressed in the present study.

Beyond the theoretical interest, solving a 2-D confinement problem can be useful in practice, mainly for quantum wires [37,38], highly oblate or flat 3D quantum dots [39,40], and 2-D quantum dots [41,42]. In the first case, quantum confinement occurs along the transverse directions of the wire, which is where the 2-D character of the SE comes from [43]. The 2-D SE energy solutions under the effective mass approximation give the subband edges in the quantum wires. In the second case, the 3-D SE specific to a quantum dot can be adiabatically decoupled into a 1-D problem of strong confinement along the small size of the nanostructure and a transverse 2-D problem with a modified potential [44]. In the third case, calculations of the electronic properties are usually performed by expressing the Hamiltonian in a basis function set of atomic orbitals or by using the density functional theory [45,46].

In this study, we propose a deep NN with two hidden layers (HLs) and thousands of subnets to estimate the ground state energy and wave function of a particle confined in an arbitrary 2-D QW. The NN was trained using a set of CPs, energies, and WFs previously generated using the finite element method (FEM). The NN can be understood as a set of separately trained subnets for each element of the position discretization. This makes the training process more transparent and allows for parallelization. Several accuracy indicators have been proposed for the NN testing. The subnets are trained on a large dataset (DS) using the stochastic gradient descent (SGD) method with variable data batches, and the training is validated with respect to a second similar DS. The network was then tested with a third DS, prepared using a different algorithm. In addition, several cases of analytical CP have been solved and discussed.

The contents of the work are as follows: Section 2 contains the statement of the problem and the mathematical principles of our approach; Section 3 presents in detail the obtaining of data samples, the training of the NN, the validation of the results obtained, and their testing based on several accuracy indicators; general conclusions are given in Section 4; examples and technical details concerning the potential samples of the DSs are provided in the Appendix.

## 2. Methods

### 2.1. Two-dimensional Schrödinger equation and sample data

If a constant effective mass approach is used, the time-independent SE of a particle confined in a 2-D QW is

$$-\frac{\hbar^2}{2m^*}\left[\frac{\partial^2 \varphi(x,y)}{\partial x^2} + \frac{\partial^2 \varphi(x,y)}{\partial y^2}\right] + V(x,y)\varphi(x,y) = E\varphi(x,y). \quad (1)$$

The potential energy

$$V(x,y) = \begin{cases} V_i(x,y), & \sqrt{x^2 + y^2} < R_o \\ V_o, & \sqrt{x^2 + y^2} \geq R_o \end{cases} \quad (2)$$

is defined such that $0 \leq V_i(x,y) \leq V_o$ and the discontinuity domain of $V_i(x,y)$ has a Lebesgue measure of zero. $V_o$ is the maximum "depth" of the QW, and $R_o$ is the outermost radius of the confinement zone, that is, of the circle including the subdomain of $\mathbb{R}^2$ where $V_i(x,y) < V_o$.

We denote by $R$ the "effective radius" of the QW, defined as the radius of the circle circumscribing the same area as the confinement zone. With the notations $\xi \equiv x/R$, $\eta \equiv y/R$, $\psi(\xi,\eta) \equiv \varphi(R\xi, R\eta)$, $v(\xi,\eta) \equiv$ $V(R\xi, R\eta)/V_o$, $e \equiv E/V_o$, and the dimensionless scale factor $\mu \equiv 2m^*V_oR^2/\hbar^2$, the Eq. (1) can be expressed in the following standardized form:

$$-\frac{1}{\mu}\left[\frac{\partial^2 \psi(\xi,\eta)}{\partial \xi^2} + \frac{\partial^2 \psi(\xi,\eta)}{\partial \eta^2}\right] + v(\xi,\eta)\psi(\xi,\eta) = e\psi(\xi,\eta), \quad (3)$$

where $0 \leq v(\xi,\eta) \leq 1$ and $e \in (0,1)$. The dimensionless quantity $e$ will continue to be referred to as energy. The real WF can be range-normalized so that $max|\psi(\xi,\eta)| = 1$.

We denote by $\{v_\sigma\}_{1\leq\sigma\leq S}: \mathbb{R}^2 \to [0,1]$ a set of $S$ dimensionless confinement functions defined in such a way as to ensure the existence of the ground bound state $\{\psi_\sigma, e_\sigma\}$ for each of them. In any case of a bound state in the QW, the WF decreases exponentially towards zero outside the geometric confinement domain such that $\lim_{\rho(\xi,\eta)\to\infty} \psi_\sigma = 0$, where $\rho(\xi,\eta) \equiv \sqrt{\xi^2 + \eta^2}$ is the radial position. An FEM numerical solver using the Dirichlet boundary condition at $\rho(\xi,\eta) = r_b$ may be used to approximate the ground state WFs $\{\psi_\sigma\}_{1\leq\sigma\leq S}$ and energies $\{e_\sigma\}_{1\leq\sigma\leq S}$. $r_b \equiv R_b/R$ denotes the circular boundary radius, which is sufficiently high with respect to $r_o \equiv R_o/R$ such that the calculation accuracy is satisfactory. The FEM uses a mesh of nodes $\Xi_N \equiv \{(\xi_n, \eta_n)\}_{1\leq n\leq N}$ densely distributed inside a circle of radius $r_b$. In the following, we use the notation $\psi_\sigma(\Xi_N) \equiv [\psi_\sigma(\xi_n, \eta_n)]^t_{1\leq n\leq N} \equiv [\psi_\sigma^n]^t_{1\leq n\leq N}$ for the column vector of the FEM numerical solution corresponding to the sample CP $v_\sigma(\Xi_N) \equiv [v_\sigma(\xi_n, \eta_n)]^t_{1\leq n\leq N} \equiv [v_\sigma^n]^t_{1\leq n\leq N}$. We call "sample-problem", or "sample" for short, the set consisting of a CP function $v_\sigma$, the WF of the ground level $\psi_\sigma$, and the corresponding energy of the particle $e_\sigma$, as they are calculated by the FEM. A collection of samples is called a "dataset". At least two different DSs $\{v_\sigma; \psi_\sigma; e_\sigma\}_{1\leq\sigma\leq S}$ defined in this way are necessary, one for training and validating the NN, and the other for testing it.

### 2.2. Neural networks architecture and underlying functions

Since the mesh used by the FEM can be extremely fine, the data sampling used for the NN may be done in a subset $\Xi_M \equiv \{(\xi_i, \eta_i)\}_{1\leq i\leq M}$ of the mesh, provided that it sufficiently covers the bounded domain on which the standardized SE is defined. The same mesh subset was used for sampling the input data (values of the CP functions) and estimating/predicting the output data (values of the ground state WFs). A deep NN with two HLs is proposed, as shown in Fig. 1. The mesh subset determines the number $M$ of neural nodes in both the input and output layers. Figure 1(a) shows that the NN can be decomposed into $M$ similar separate subnets, which allows easy parallelization of the calculations. Each subnet receives the same data input from all nodes in the input layer (IL) and has a single output node, which is the estimate of the WF at a single mesh point. The subnets are identical in their internal structure but differ in functionality: their neurons are not equivalent after training the network. Figure 1(b) shows the neural architecture of a single subnet enclosing two HLs, each with $P$ neurons. In a subnet, all nodes of a layer are interconnected with all the nodes of the previous layer, and each neural connection is mathematically coded using a weight coefficient. The number of neural connections in a subnet is $P(M + P + 1)$ such that the total number of weights in the NN is $MP(M + P + 1)$.

The activation function of the neurons of the output layer (OL) is the widely used standard logistic sigmoid (sigm), whose codomain fits the interval into which the values of the range-normalized WFs fall. The neurons of the HLs have the hyperbolic tangent (tanh) as an activation function. We selected these related functions because they are continuously differentiable in $\mathbb{R}$ and have simple derivatives.

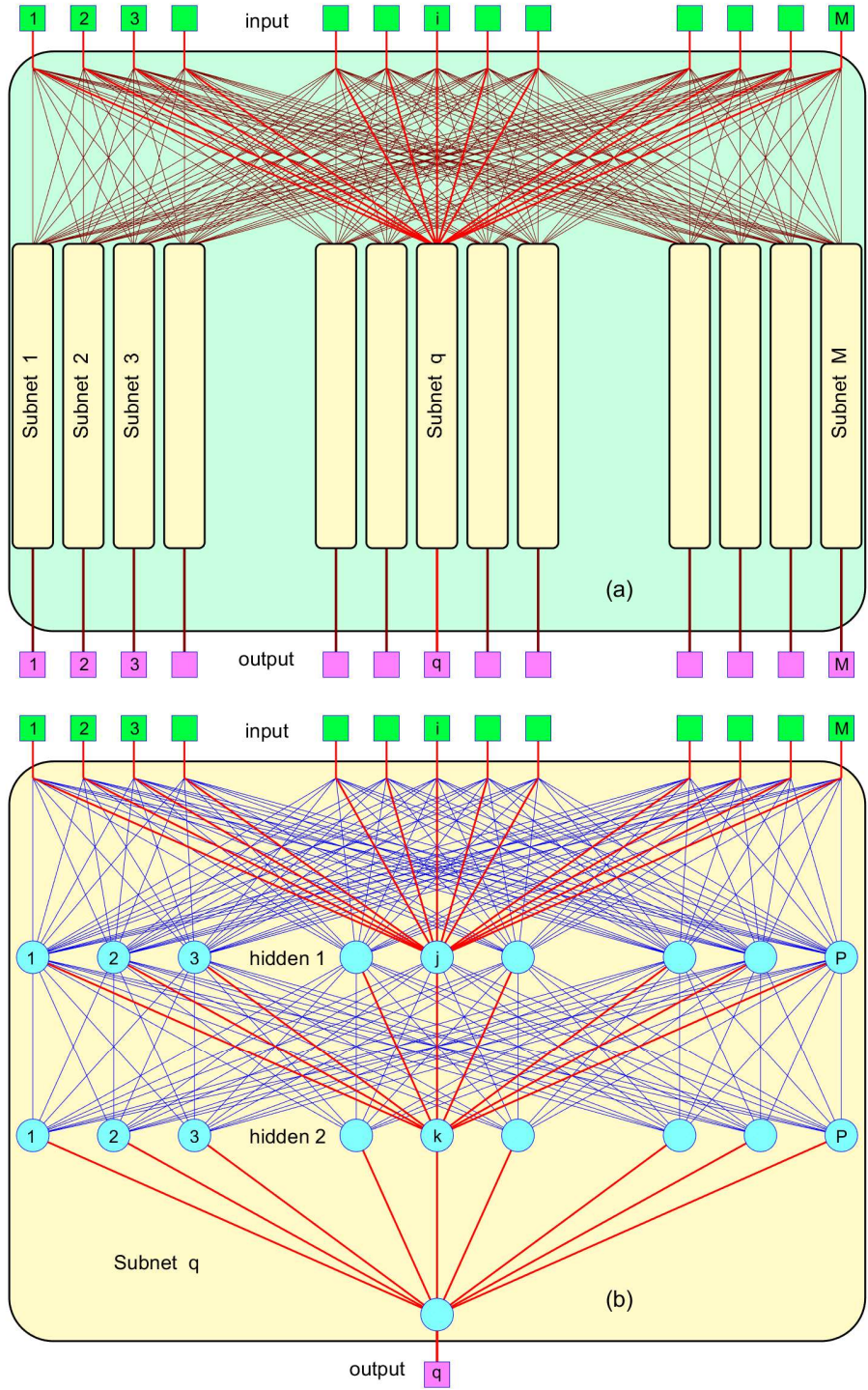

**Figure 1.** (a) Neural network composed of *M* independent subnets; both input and output layers have *M* nodes; (b) Subnet with *M* nodes in the input layer, *P* nodes in each of the hidden layers, and one output node.

Because the functioning of the entire NN can be reduced to the mathematics of its subnets, we explain the flow of data in a single generic subnet q. In the following expressions, all matrix operations are element-wise, except for the matrix product explicitly denoted by "·".

Given a particular CP function $v_\sigma$, the data sent by HL1 to a neuron in HL2 are

$$(h_1)_\sigma^q = H_1 \tanh[\Lambda_1^q \cdot v_\sigma(\Xi_M)] \equiv H_1 \frac{\exp[2\Lambda_1^q \cdot v_\sigma(\Xi_M)]-1}{\exp[2\Lambda_1^q \cdot v_\sigma(\Xi_M)]+1}, \quad (4)$$

where $(h_1)_\sigma^q$ is a $P \times 1$ column vector, $H_1$ is a scale coefficient, $\Lambda_1^q$ is the $P \times M$ weight matrix of HL1 of subnet q, and $v_\sigma(\Xi_M)$ denotes the $M \times 1$ column vector of the sample CP data in the IL, that is, $[v_\sigma(\xi_i, \eta_i)]_{1 \leq i \leq M}^t \equiv [v_\sigma^i]_{1 \leq i \leq M}^t$.

The data sent by the HL2 to the output neuron is

$$(h_2)_\sigma^q = H_2 \tanh[\Lambda_2^q \cdot (h_1)_\sigma^q] \equiv H_2 \frac{\exp[2\Lambda_2^q \cdot (h_1)_\sigma^q]-1}{\exp[2\Lambda_2^q \cdot (h_1)_\sigma^q]+1}, \quad (5)$$

where $(h_2)_\sigma^q$ is a $P \times 1$ column vector, $H_2$ is a scale coefficient, and $\Lambda_2^q$ is the $P \times P$ weight matrix of HL2 of the subnet q.

The estimated WF value at node q of the submesh $\Xi_M$, for the CP $v_\sigma$, that is, the subunitary output of the subnet q, is

$$\tilde{\psi}_\sigma(\xi_q, \eta_q) \equiv \tilde{\psi}_\sigma^q = \text{sigm}[\Lambda_o^q \cdot (h_2)_\sigma^q] \equiv \frac{1}{1+\exp[-\Lambda_o^q \cdot (h_2)_\sigma^q]}, \quad (6)$$

where $\Lambda_o^q$ is the $1 \times P$ weight-row vector of the OL of the subnet q.

Combining Eqs. (4-6) into a compact expression, we get

$$\tilde{\psi}_\sigma^q = \text{sigm}[H_2 \Lambda_o^q \cdot \tanh[H_1 \Lambda_2^q \cdot \tanh[\Lambda_1^q \cdot v_\sigma(\Xi_M)]]]. \quad (7a)$$

An extra single subnet can be used for ground-level energy estimation. Formally replacing $\tilde{\psi}_\sigma^q$ with $\tilde{e}_\sigma$ and index q by e, we obtain

$$\tilde{e}_\sigma = \text{sigm}[H_2 \Lambda_o^e \cdot \tanh[H_1 \Lambda_2^e \cdot \tanh[\Lambda_1^e \cdot v_\sigma(\Xi_M)]]]. \quad (7b)$$

Further we will use the notation $\tilde{\psi}_\sigma(\Xi_M) \equiv [\tilde{\psi}_\sigma^q]_{1 \leq q \leq M}^t$, i.e. the neural estimation of the FEM solution $\psi_\sigma(\Xi_N)$.

### 2.3. Loss function, weight optimization and network training

Training the NN involves approaching the optimal values of the weight matrices such that the network estimates are as close as possible to the solutions expected from the training DS: that is, the loss function is minimized. Optimization is performed iteratively using the gradient descent (GD) method, starting from the initial values $\{\Lambda_1^q(0); \Lambda_2^q(0); \Lambda_o^q(0)\}, q \in \{1,2,\dots,M\}$. The updated weight matrices and the neural estimation of the solution after τ iterations will be denoted by $\{\Lambda_1^q(\tau); \Lambda_2^q(\tau); \Lambda_o^q(\tau)\}, q \in \{1,2,\dots,M\}$ and $[\tilde{\psi}_\sigma(\Xi_M)](\tau)$, respectively.

In this study, the global NN loss function corresponding to the training DS is defined as

$$\mathcal{L}(\tau) \equiv \sum_{\sigma=1}^{S} \left\| [\tilde{\psi}_\sigma(\Xi_M)](\tau) - \psi_\sigma(\Xi_M) \right\|^2 \equiv \sum_{q=1}^{M} L^q(\tau), \quad (8)$$

where $\|\cdot\|$ is the Euclidian norm on $\mathbb{R}^M$ and $L^q(\tau)$ is the local subnet q loss function:

$$L^q(\tau) = \sum_{\sigma=1}^{S} [\tilde{\psi}_\sigma^q(\tau) - \psi_\sigma^q]^2. \quad (9a)$$

Additionally, the energy loss function can be defined in a similar way

$$L^e(\tau) = \sum_{\sigma=1}^{S} [\tilde{e}_\sigma(\tau) - e_\sigma]^2, \quad (9b)$$

where $\tilde{e}_\sigma(\tau)$ and $e_\sigma$ are the neural estimations of the energy after $\tau$ iterations and the expected energy, respectively.

Because each subnet is trained independently, the loss function can be minimized separately for each output node q, as presented below. The gradient components of the loss function $L^q(\tau)$ with respect to the weights are the $P \times M$, $P \times P$, and $1 \times P$ matrices, respectively:

$$\nabla L^q(\tau) \equiv \left\{ \frac{\partial L^q}{\partial \Lambda_1^q}(\tau); \frac{\partial L^q}{\partial \Lambda_2^q}(\tau); \frac{\partial L^q}{\partial \Lambda_0^q}(\tau) \right\}. \tag{10a}$$

The gradient of the energy loss function $L^e(\tau)$ is

$$\nabla L^e(\tau) \equiv \left\{ \frac{\partial L^e}{\partial \Lambda_1^e}(\tau); \frac{\partial L^e}{\partial \Lambda_2^e}(\tau); \frac{\partial L^e}{\partial \Lambda_0^e}(\tau) \right\}. \tag{10b}$$

The starting values of all the weight coefficients are randomly chosen and determine the initial values $L^q(0)$ and $L^e(0)$ of the subnet loss functions. The weight matrices and, implicitly, the loss functions are iteratively updated by a first-order approximation to minimize the losses in Eqs. (9a-9b), that is, the GD method:

$$\{\Lambda_1^q(\tau+1); \Lambda_2^q(\tau+1); \Lambda_0^q(\tau+1)\} = \{\Lambda_1^q(\tau); \Lambda_2^q(\tau); \Lambda_0^q(\tau)\} - \lambda \nabla L^q(\tau), \tag{11a}$$

$$\{\Lambda_1^e(\tau+1); \Lambda_2^e(\tau+1); \Lambda_0^e(\tau+1)\} = \{\Lambda_1^e(\tau); \Lambda_2^e(\tau); \Lambda_0^e(\tau)\} - \lambda \nabla L^e(\tau). \tag{11b}$$

Here, $\lambda$ is the learning rate and $\tau$ is an integer index such that $0 \leq \tau < T$, where $T$ denotes the maximum number of iterations. Theoretically, if the loss function decreases monotonically, the value of $T$ can be established based on the criterion imposed by the maximum allowed variation in the loss function from one iteration to another. Because working with DSs of hundreds of thousands of samples is, in practice, a very expensive computational burden, one may opt for the SGD variant of the GD method. This method replaces the gradient calculation based on the complete DS containing $S$ samples by an estimate calculated from a randomly selected batch of $S'$ samples ($S' \ll S$) which can be totally or partially changed at each iteration. The SGD method achieves faster iterations; however, the convergence has a lower rate and fluctuating behavior. Because the loss function has fluctuations that overlap with the average decreasing trend, $T$ should be imposed by the average behavior of the loss function and/or by the available computing resources.

### 2.4. Testing the network. Accuracy indicators

Several quantitative indicators for testing the trained NNs are proposed based on the dispersion from the expected values of the WF, predicted energy, and average position of the particle. The indicators are calculated for each sample $\sigma$ in a DS ($0 \leq \sigma \leq S$). Network efficiency is determined by analyzing and comparing the distribution of the values of these indicators in the DSs involved.

The main accuracy indicator is the relative difference between the NN-estimated WF $\tilde{\psi}_\sigma$ and the FEM-calculated solution $\psi_\sigma$, that is, the WF relative deviation

$$\epsilon_\sigma \equiv \frac{\|\tilde{\psi}_\sigma(\Xi_M) - \psi_\sigma(\Xi_M)\|}{\|\psi_\sigma(\Xi_M)\|}. \tag{12}$$

The spatial overlap of the exact and estimated WFs may be another indicator of the NN accuracy. The estimated WF relative spatial overlap is defined as

$$\omega_\sigma \equiv \frac{\sum_{q=1}^{M} \tilde{\psi}_\sigma^q \psi_\sigma^q}{\|\tilde{\psi}_\sigma(\Xi_M)\| \|\psi_\sigma(\Xi_M)\|}. \tag{13}$$

The NN-estimated average positions and FEM-calculated average positions are compared by calculating the deviations of the average $\xi$ and $\eta$ positions, respectively:

$$\Delta\langle\xi\rangle_\sigma \equiv \frac{\sum_{q=1}^{M} \xi_q (\tilde{\psi}_\sigma^q)^2}{\|\tilde{\psi}_\sigma(\Xi_M)\|^2} - \frac{\sum_{q=1}^{M} \xi_q (\psi_\sigma^q)^2}{\|\psi_\sigma(\Xi_M)\|^2}, \tag{14a}$$

$$\Delta\langle\eta\rangle_\sigma \equiv \frac{\sum_{q=1}^{M} \eta_q (\tilde{\psi}_\sigma^q)^2}{\|\tilde{\psi}_\sigma(\Xi_M)\|^2} - \frac{\sum_{q=1}^{M} \eta_q (\psi_\sigma^q)^2}{\|\psi_\sigma(\Xi_M)\|^2}. \tag{14b}$$

As the estimated WF approaches the exact WF, the limits of $\varepsilon_\sigma$, $\omega_\sigma$, $\Delta\langle\xi\rangle_\sigma$ and $\Delta\langle\eta\rangle_\sigma$ are 0, 1, 0 and 0, respectively.

The deviation of the NN-estimated energy with respect to the FEM-calculated value is:

$$\Delta e_\sigma \equiv \tilde{e}_\sigma - e_\sigma \tag{15a}$$

or, in a relative expression:

$$\kappa_\sigma \equiv \frac{\tilde{e}_\sigma}{e_\sigma} - 1. \tag{15b}$$

## 3. Calculations and Results

In the following, we use the abbreviation ksamp (kilosample) to denote 1000 samples. To train, validate, and test the NN, three DSs have been prepared (hereinafter referred to as DS1, DS2, and DS3), each of them containing $S = 100$ ksamp. DS1 is used for training, DS2 for validation, and DS3 is reserved for testing. Various algorithms have been implemented to generate arbitrary CPs for DSs, as explained in the Appendix. DS1 and DS2 are distinct but they were prepared using the same randomization method, whereas DS3 was prepared using a different algorithm. When we refer to the output provided by the NN, we usually use the word "estimate" but if we want to emphasize that the input was not the training DS, we may use the word "prediction" instead.

### 3.1. Two-dimensional finite element calculation and sample preparation

COMSOL Multiphysics® FEM software was used to generate samples [47]. To compute the expected ground state WFs and energies, the FEM model was built to solve the SE with $r_b = 5$ and $\mu = 26$ for all CP functions in the DSs. A very small value of the border radius ($r_b < 2r_o$) may lead to an artificial increase in the confinement, whereas an excessively large value (under the conditions of a prefixed number of nodes) will lead to an important decrease in the density of nodes in the confinement area, which will further negatively influence the accuracy of the calculation. The reference scale factor $\mu$ was reasonably chosen in accordance with the typical values of real confining systems, so that several bound energy levels exist for all CPs [37,48]. It is noteworthy that the scale factor introduced in Eq. (3) is proportional to what we could call the "confining volume" $\pi V_o R^2$, that is, a cylindrical pseudovolume with the real confining base area $\pi R^2$ and the energetic "height" $V_o$. Intuitively, the NN is trained on a set of problems with confining volumes distributed around the chosen reference value. The variance is introduced through the very algorithm that generates the CPs of the DS by randomly changing the confining perimeter and the variations in the potential inside the confining zone. The numerical choice of $\mu$ therefore does not mean drastic particularization, but rather offers a plausible reference value in relation to the physical systems of interest. For example, for cylindrical confinement in semiconductor wires based on GaAs/AlGaAs, the value of $\mu$ previously defined is given by a realistic confinement radius of approximately 8 nm. For any given material, a scale factor that is too small corresponds to an extreme confinement regime, which is difficult to achieve in real semiconductor structures. In addition, for the large values of the kinetic term of the Hamiltonian that are reached in very small semiconductor structures, the effective mass approximation is questionable [2]. Too high values of $\mu$ correspond to systems with a high density of energy levels, in which the quantization fades and the interest in solving the SE is limited. We define a user-controlled mesh of nodes $\Xi_N \equiv \{(\xi_n, \eta_n)\}_{1 \leq n \leq N}$, unevenly distributed inside a circle of radius $r_b$. Nodes are relatively rare close to the boundary, densely distributed in the vicinity of the QW perimeter, and very dense in the confining zone. When solving a 2-D differential problem, the FEM software considers the standard physics-controlled mesh with approximately 13000 nodes to be "extremely fine" [47]. Indeed, for typical CPs, the spatial element size of this mesh is sufficiently small such that the accuracy of solving the SE is sufficiently high. However, in this study, we allowed the random CPs of the DSs to exhibit fast variations and sometimes several points or lines of discontinuity. Therefore, we chose a user-controlled mesh with a larger total number of nodes, that is $N = 18724$. Given that for any bound state, the WF decreases exponentially to zero sufficiently far from the confinement zone, the Dirichlet boundary condition $\psi(r_b) = 0$ may be assumed. We determined that the mesh was sufficiently refined and the chosen value of the parameter $r_b$ was sufficiently high to ensure an accuracy better than $5 \times 10^{-5}$ for the calculation of the ground state energy. The typical error can be estimated by comparing the FEM result $e = 0.154021$ with the exact semi-analytical solution $e_{sa} = 0.154005$, in the particular case of a finite-wall cylindrical confinement with $v(\xi, \eta) = H(\rho(\xi, \eta) - 1)$, where $H$ denotes the Heaviside step function [49].

Figure 2 shows the graphical details of the mesh used, at various degrees of image magnification. The mesh is triangular, has circular symmetry and contains four distinct concentric zones colored blue in the figures. The outer area (we call it the "far mesh" in Fig. 2a), has a low density of nodes, because in this region the values of the WFs are very small and slowly variable. This is the circular crown between the circle of radius $r_3 = 1.3\sqrt{2\pi/3/\sin(2\pi/3)} \cong 2$ and the border of radius $r_b$. Its inner radius $r_3$ was calculated such that none of the irregular confinement zones generated by the CP randomization algorithms would enter this zone ($r_o < r_3$ for all samples). Another region in the shape of a circular crown of inner radius $r_2 = 1.2\sqrt{2\pi/4/\sin(2\pi/4)} \cong 1.5$ and outer radius $r_3$ follows inwards (the "intermediate mesh" in Fig. 2b). Its node density is higher than that of the outer mesh but remains relatively low. This mesh domain ensures the adaptation between the low-density outer mesh and the high-refinement mesh of the confining perimeter, and only a small number of samples with a highly eccentric confining perimeter marginally penetrate this zone. The next smaller circular crown of the mesh (the "near mesh" in Fig. 2c) is highly refined and corresponds to the region where the CP can already exhibit large variations from the external constant value to lower values; that is, it contains large portions of the confining perimeter. The main zone of the mesh (the "central mesh" in Fig. 2d), shaped like a circular disk with radius $r_1 = 1.1$, is extremely refined and corresponds to the confinement zone in which the most important variations in the CP and WF occur. Only a small number of samples (those with reduced eccentricity, close to a circular confinement perimeter) had confining zones that were completely contained in the central mesh domain ($r_o < r_1$).

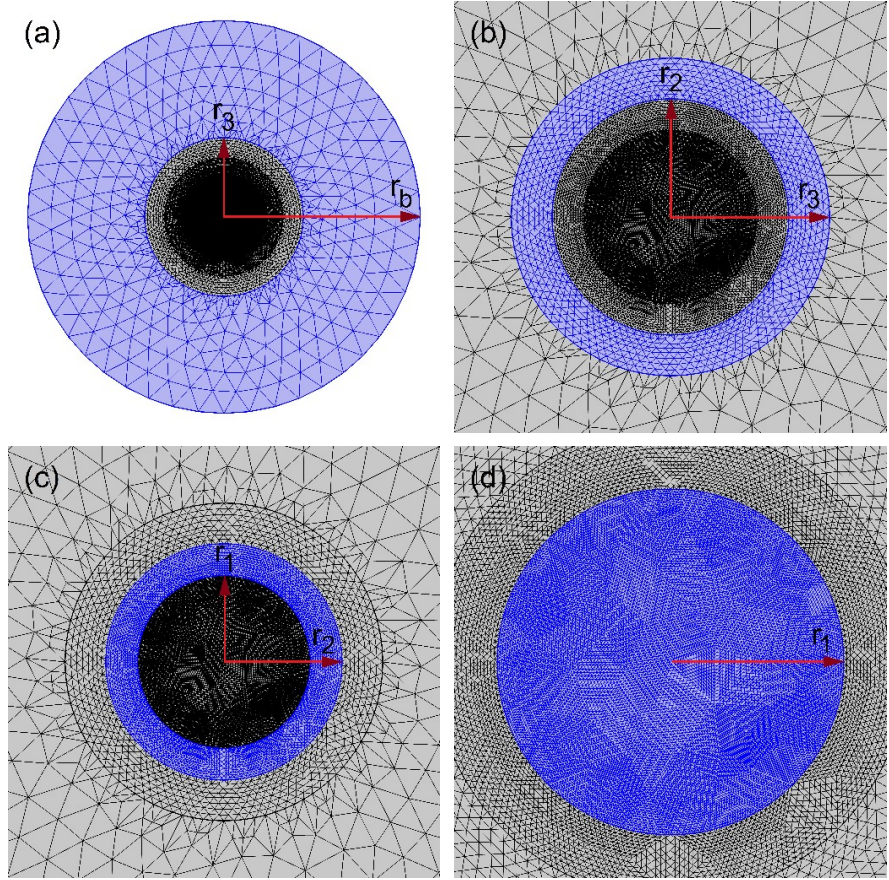

**Figure 2.** Mesh domains: (a) Far mesh containing about 2.4% of the total number of nodes; (b) Intermediate mesh containing about 5.1% of nodes; (c) Near mesh containing approximately 18.5% of nodes; (d) Central mesh consisting of approximately 77.6% of nodes.

The complete technical details of the mesh used are presented in Table I.

**Table I.** Concentric domains of the finite element method mesh and their main geometric characteristics

| Mesh entity | Radial position | Mesh nodes (vertices) | Percentage from total number of nodes | Edge elements | Edge length | Average element length | Element length ratio |
|---|---|---|---|---|---|---|---|
| | | | | Triangular elements | Mesh area | Average element area | Element area ratio |
| *Boundary* | $\rho = r_b$ | *48* | *0.256%* | *48* | *31.39* | *0.654* | *1* |
| Domain (a) | $\rho \in [r_3, r_b]$ | 451 | 2.409% | 758 | 65.48 | 0.0864 | 0.0544 |
| *Border (a-b)* | $\rho = r_3$ | *96* | *0.513%* | *96* | *12.7* | *0.1323* | *0.9828* |
| Domain (b) | $\rho \in [r_2, r_3]$ | 960 | 5.127% | 1632 | 5.726 | 35.09E-4 | 0.2309 |
| *Border (b-c)* | $\rho = r_2$ | *192* | *1.025%* | *192* | *9.449* | *0.0492* | *0.9821* |
| Domain (c) | $\rho \in [r_1, r_2]$ | 3456 | 18.458% | 6336 | 3.304 | 5.215E-4 | 0.2079 |
| *Border (c-d)* | $\rho = r_1$ | *384* | *2.051%* | *384* | *6.911* | *0.018* | *0.9387* |
| Domain (d) | $\rho \in [0, r_1]$ | 14529 | 77.596% | 28672 | 3.801 | 1.326E-4 | 0.3224 |
| Total | $\rho \in [0, r_b]$ | $N = 18724$ | 100% | 720 / 37398 | 78.32 | 2.094E-3 | 3.603E-4 |

SE was solved for all $3 \times 10^5$ randomized CPs in the training, validation, and testing sets. The energies $e_\sigma$ and WFs $\psi_\sigma$ of the ground level were stored, forming together with the corresponding CPs $v_\sigma$ what we call the "data sets" DS1, DS2 and DS3. In the following, we consider that these energy and WF solutions represent the true (exact) values. Figure 3 shows the energies obtained for each of the three sets of random CPs. The mean energy value $e_m$ over the entire set, the most frequent (probable) energy $e_p$, and the standard energy deviation $\delta e$ are indicated. The color scale illustrates the deviation of the energy of each sample from the mean value of the set. Histograms of the energy occurrence frequencies were created by dividing the energy interval (0,1) into 100 equal bins and counting the results in each bin. The numbers on the horizontal axis of the histograms also indicate the linear probability density.

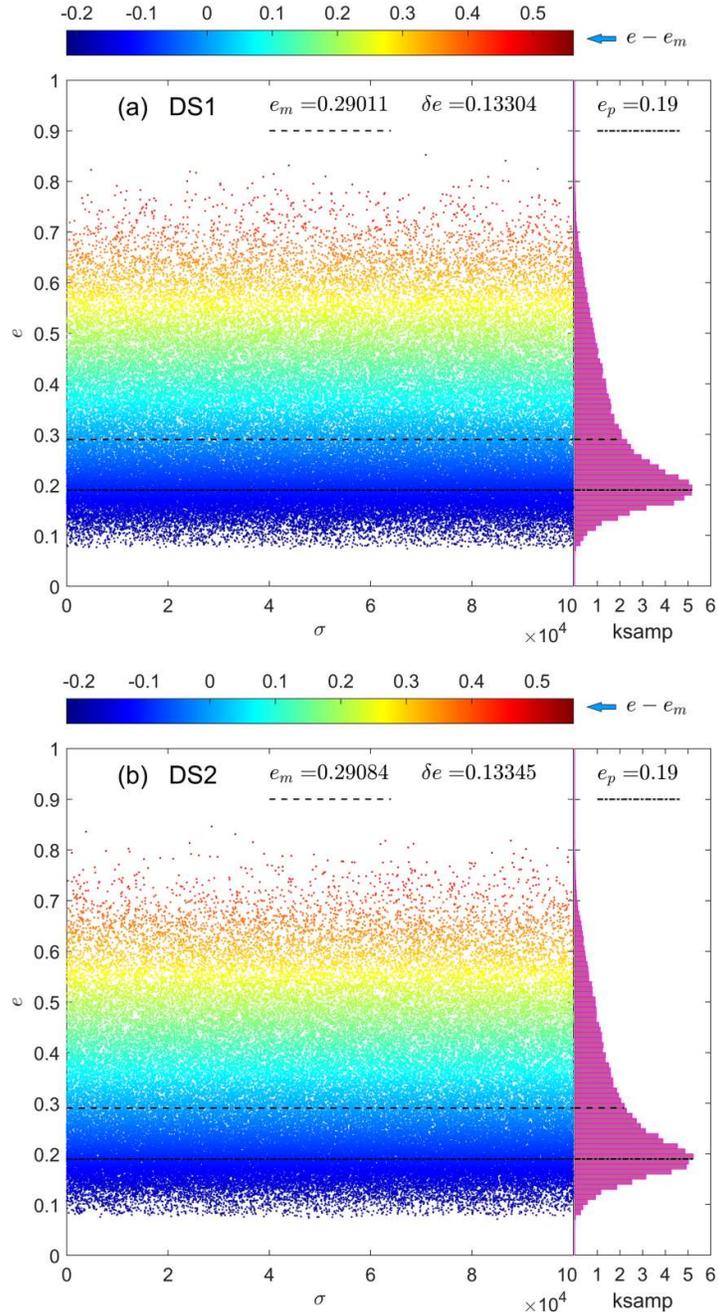

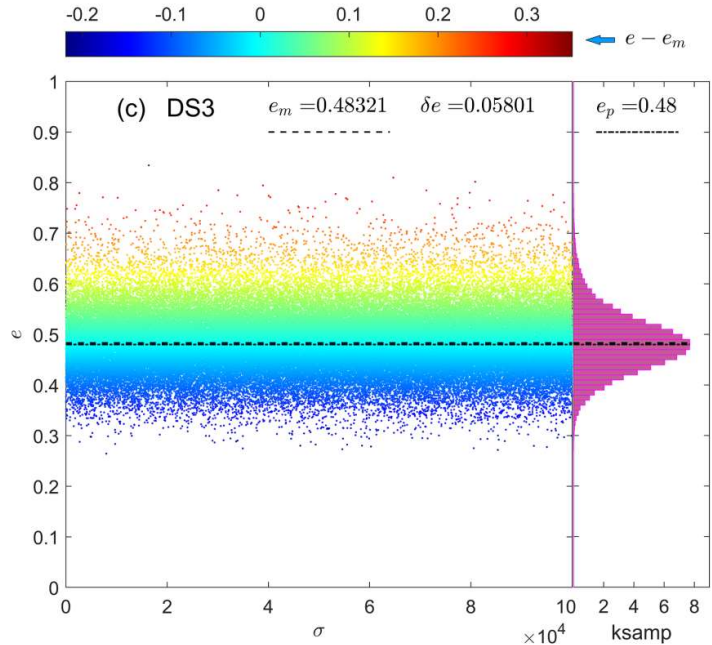

**Figure 3.** Energy distribution of the ground level corresponding to the sets of confinement potentials: a) DS1 – training set; b) DS2 – validation set; c) DS3 – testing set. The lateral histograms illustrate the occurrence frequency of the results as a function of energy.

As expected, the energy distributions obtained for DS1 and DS2 are very similar (Figs. 3a and 3b, respectively) and confirm that the sets are sufficiently large to be representative of their common CP randomization algorithm. In these cases, the upper positive skewness of the frequency histograms indicates that the energies are unevenly distributed around the mean. Thus, it is more likely to obtain energies higher than the most frequent value. Figure 3a shows the results obtained with the CPs of the third set generated using a different randomization algorithm. The mean energy value was significantly higher and almost equal to the most frequent value. The appearance of the distribution is quite different, almost symmetrical, and shows aspects of a normal distribution. The standard deviation of the energy was also much smaller than that in the previous cases.

### 3.2. Defining the input/output layers of the neural network

The SE was solved numerically using a mesh with a relatively large number of nodes, which was justified by the need to obtain reliable results in cases with rapid variations in the CP or many discontinuities. However, it is impractical to associate each computational node with an input/output node in the NN because the ground state WF generally exhibits a slow variation with position. Nevertheless, if computational cost had not been a limiting factor, we would have opted for a full representation of the FEM mesh in the IL and OL of the NN. We are aware that better effectiveness of the NN would be obtained if the IL corresponded to an even denser representation of the nodes in the original mesh. In this work, we chose a subset $\Xi_M \equiv \{(\xi_i, \eta_i)\}_{1 \leq i \leq M}$ of the FEM calculation mesh to represent both the IL and OL of the NN. The manner in which we selected the nodes is illustrated in Fig. 4 and is based on the intention of approximately uniform coverage of a circular region that is representative of the most important variations of the CP and WF. The disk containing all the nodes of the selection was chosen to be of radius $r_a = 1.75$, which is intermediate between $r_2$ and $r_3$ (Fig. 4a). Because this domain contains regions with various mesh refinements and the original mesh is not regular, we selected the original nodes that are closest to the vertices of the triangular tiling {3,6} (Fig. 4b). These are the centers of the densest possible circle packing in the plane. The number of nodes in the subset is then controlled by a single parameter, that is, tiling edge length. If we choose this parameter to be 0.06, we get $M = 2764$. The CP values in these nodes represent the input data of the NN that estimates the WF as well as the input of the subnet that estimates the energy of the ground level.

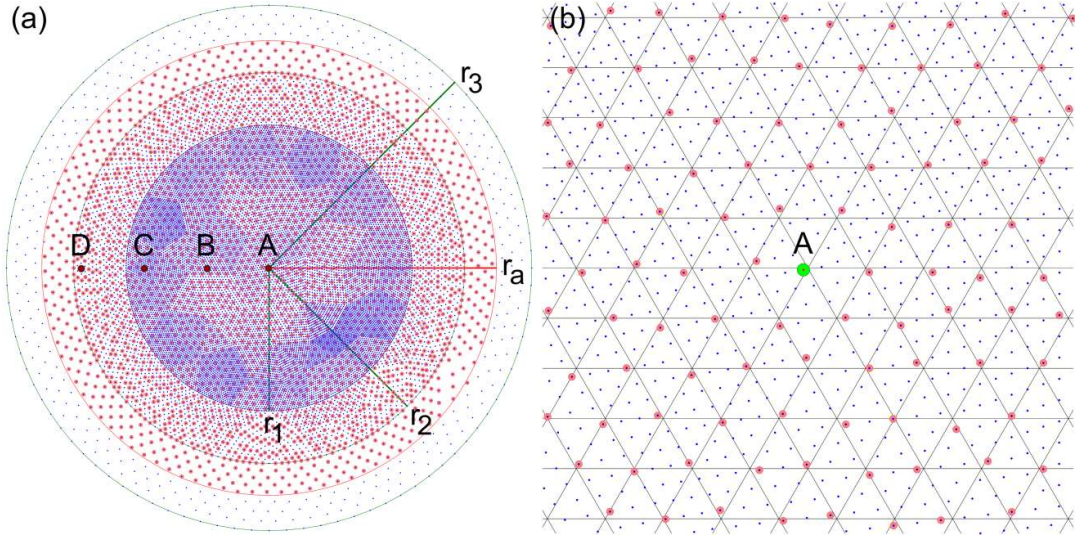

**Figure 4.** a) The mesh subset $\Xi_M \equiv \{(\xi_i, \eta_i)\}_{1 \leq i \leq M}$ intended for neural network training: blue dots are the original mesh nodes of the finite element method and red dots are the $M = 2764$ selected nodes. Several particular nodes marked A, B, C, D were chosen to further illustrate the training of the corresponding neural subnets and the distribution of the wave function estimation errors. b) Details of the triangular tiling and the selected nodes in the vicinity of the central node A.

### 3.3. Subnet training and energy evaluation

Heuristically optimizing the NN architecture is prohibitively difficult because of the large time required for multiple calculations with different combinations of node numbers. Concerning neurons in the HLs, it seems that there is no strict or clear rule in the literature for setting the optimal number. Thus, we followed the empirical choice of taking it close to the geometric mean of the numbers of nodes in the IL and OL. Therefore, all subnets in this study (energy subnet included) have $M = 2764$ nodes in the IL and $P = 53$ neurons in each of the two HLs. The training of all subnets is done with the SGD method, using batches with $S' = 2$ ksamp. For this, the training DS1 of 100 ksamp random CPs was divided into subsets of 1 ksamp and, for each iteration, the current batch is rebuilt from two different subsets chosen randomly and independently. Figure 5 shows the SGD learning graph for the energy neural subnet. Because there was only one subnet for energy estimation, we were able to perform a longer training, with $T = 10^5$ iterations. However, for all other 2764 subnets used to estimate the WF, the training had to be limited to $T = 2000$ for computational cost reasons. The relatively small fluctuations that appear along the learning graph are specific to the SGD method, and are caused by the stochastic variance of the different batches used for each iteration.

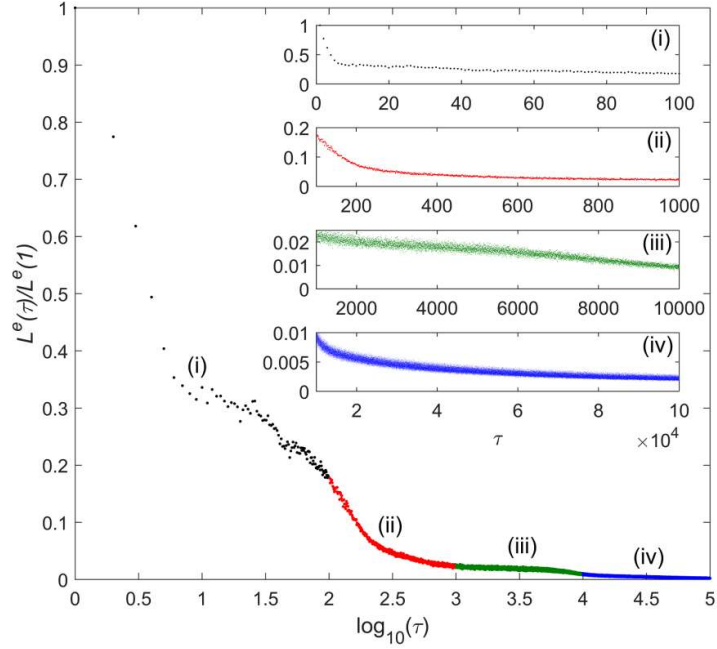

**Figure 5.** Discrete learning graph of the energy subnet: relative loss versus the number of iterations. Insets detail the loss variation on different training intervals.

The initial values $\{\Lambda_1^e(0); \Lambda_2^e(0); \Lambda_0^e(0)\}$ of the weight matrices were normally distributed random numbers with a mean 0 and standard deviations $1/100$, $1/75$ and $1/50$, respectively. We noticed that a normal distribution of the initial weights ensures better stability of the preliminary learning than a uniform distribution; however, the choice of standard deviations was rather empirical. With these initial weights, the starting value $L^e(1)$ of the loss function is obtained after the first iteration and is then used as a reference to calculate the relative loss $L^e(\tau)/L^e(1)$. The learning rate $\lambda$ must be adjusted so that the training is neither too fast nor too slow. If the learning graph decreases rapidly, it can enter a regime of strong fluctuations, instability, and even divergence. If it varies too slowly, the total training duration will increase unreasonably. In our model, the learning rate is adjusted according to the number of samples in the batch. We used a pre-learning rate $\lambda_p = 5 \times 10^{-5}$ in the first stage of the training and $\lambda = 10^{-5}$ afterwards. The change in the learning rate occurs when an empirical condition on $L^e$ is met, which is achieved after the 49th iteration, as shown in Fig. 5. If the higher value $\lambda_p$ was to be kept further, for most subnets, this would lead to instabilities and sudden increases in the cost function, compromising convergence. If the pre-learning stage is omitted by using the $\lambda$ parameter from the start, the total training time necessary to achieve the same loss function decrease will be approximately 10% higher for $T = 2000$. We divided the learning graph into four distinct intervals, as shown in the insets (i)–(iv) of Fig. 5, to better follow the variation trend of the loss function. The decrease is very fast in interval (i) and in the first part of (ii), followed by reaching a first quasi-plateau at the end of interval (ii) and the beginning of (iii), after approximately 2000 iterations. Considering that the main graph is semi-logarithmic, it should be noted that the further progress is very slow: after approximately $5 \times 10^4$ iterations, a second quasi-plateau is reached, as shown in inset (iv). The loss function decreased to 2% of the initial value after 2000 iterations and to 0.2% after $10^5$ iterations.

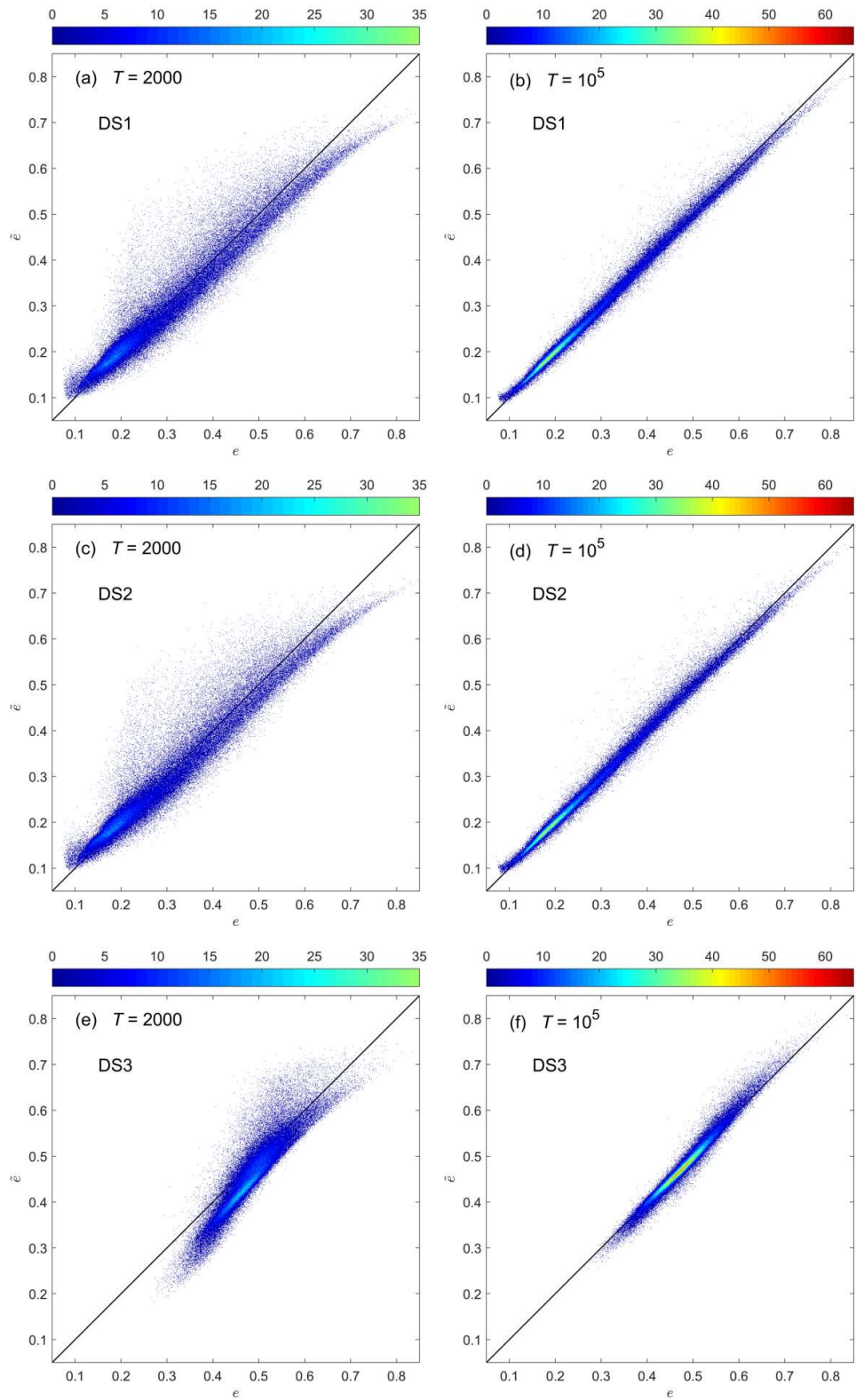

**Figure 6.** Estimated/predicted energy as a function of the true energy. The diagonal lines represent the ideal, perfect prediction. The color scale is a measure of the in-plane density of samples.

After training, the energy neural subnet was applied to the three DSs in two scenarios: T=2000 and $T = 10^5$. The results are presented in Fig. 6 in the form of bivariate histogram plots with 2-D bins. The bins were squares with side of $10^{-3}$. The true energy (FEM solution) is represented on the horizontal axis and the estimated/predicted energy (NN solution) is represented on the vertical axis. The number of samples in each bin is coded by color. The numerical labels on the color bars give the bin counts and, if multiplied by 10, provide the surface probability density. It can be observed that the distributions of the estimated energies for the training set (a,b) and validation set (c,d) are extremely similar. This is proof of the validity of the network training: that is, the number of samples in the training DS1 is large enough to ensure neural learning related to the SE itself and possibly to the algorithm generating DS1 and DS2 random CPs, but not to a particular group of samples. Figures (e) and (f) show that the predictions are also very good for DS3, which is very different from the other two. Therefore, the subnet is efficient in correctly predicting the energy of the ground state for CPs that are very dissimilar to those with which it was trained. The in-plane distribution density of the samples was significantly affected by the number of iterations used. The scattering of bins with non-zero counts is greater after only 2000 iterations (a,c,e) than after $10^5$ iterations (b,d,f). Therefore, the representative points of the graphs tended to accumulate near the diagonal lines as the subnet training improved.

It is worth mentioning that the energy can be estimated with this method without the involvement of the WF, which can be an advantage in applications where a fast response is required. For example, with the spatial discretization used in this work, the NN provides the energy approximately 60 times faster than the FEM.

### 3.4. Training and testing neural subnets for ground state WF estimation in particular nodes

To illustrate the training and predictive efficiency of the individual subnets we select 4 particular nodes of the mesh subset $\Xi_M$. These points are illustrated in Fig. 4a: A is the closest node to the origin of the coordinates, where in general the WF has relatively large values; B is approximately the midpoint of the radius of a cylindrical confinement, where there is a large dispersion of possible values; C is close to the perimeter of the confinement zone, where there are generally fast variations of the WF; and D is found outside the confinement, in a position where the values of the WFs are relatively small. There is practically no difference in the settings between the subnets used to estimate the WF and the previously described subnet for energy estimation. All $M = 2764$ subnets used for the WF estimation have the same structure and IL; they differ only in the output node. Figure 7 shows the learning graphs of the subnets related to output nodes A, B, C, and D. Subnet A learns relatively quickly, reaching a plateau value after 10 iterations, but this value of the loss function is rather high. Between iterations 100 and 2000, there was an additional decrease of approximately 5%. Based on these observations, we can anticipate that in the central area of the confinement zone, where WF maxima are usually found, underestimates of the true values are often obtained. Subnet B learns more slowly and with greater fluctuations in the relative loss precisely because the dispersion of the possible values of the WF in node B is greater than that in the other cases. After 55 iterations, there was instability in the learning graph, with a sudden increase in the loss function. In the next 50 iterations, the learning stabilizes, and it is observed that around the $200^{th}$ iteration, there is a transition to a 5 times lower value of the learning rate, after which the fluctuations remain relatively small. The learning of subnets A and B shows, by the slope of the graphs between iterations 1000 and 2000, that they will still have the potential to decrease if a larger number of iterations is practically feasible. The behavior of the learning graph of subnet C shows some notable differences compared to B: no instability occurs, it seems to reach a plateau value after 1000 iterations, and the value of the loss function after 2000 iterations reaches 5% of the initial value, which is much lower than in the previous cases. Finally, subnet D learns very quickly at the beginning, similar to A, but the learning curve decreases more. After 1000 iterations, the loss function becomes less than 1% of the initial value, and the slope seems to show that the subnet still has a slight learning potential. The partial similarity between the graphs of subnets A and D comes from the fact that, in both cases, the marginal codomain of the logistic activation functions is involved, close to 1 and 0, respectively. Learning reaches the plateau regime faster, because the activation function derivative of the output neurons decreases rapidly. After the same number of iterations, the relative decrease in the loss function exhibited the following trend from the center to the outside: ~7% for A, ~12% for B, ~5% for C, and ~0.7% for D.

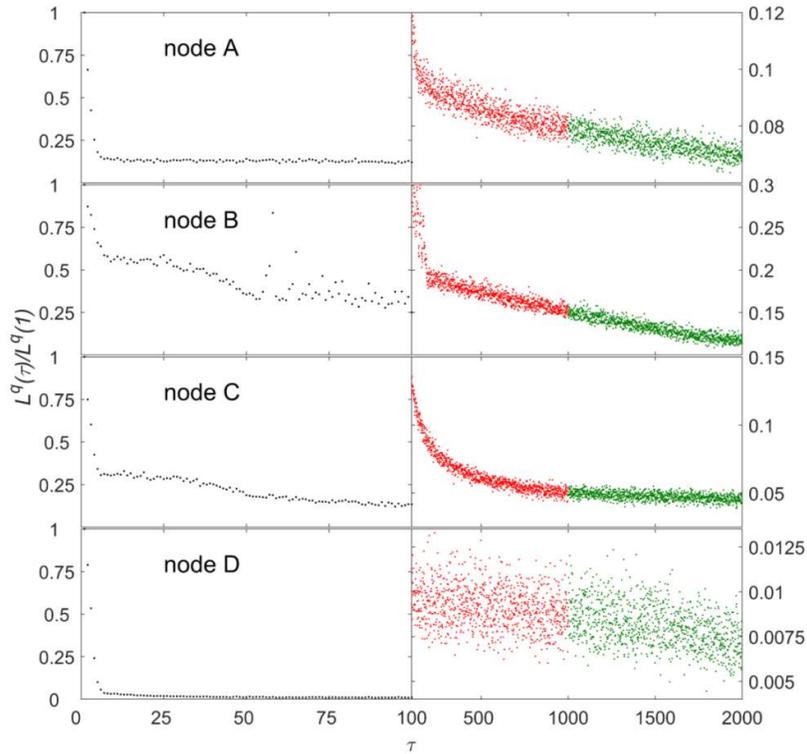

**Figure 7.** Learning graphs of the subnets corresponding to the nodes marked in Fig. 4a.

Trained neural subnets A, B, C, and D were applied to the training and testing DSs. Bivariate histograms are shown in Fig. 8. The bins are squares with side of $10^{-3}$, the true value of the WF (FEM solution) is on the horizontal axis, and the estimated/predicted value (NN solution) is on the vertical axis. The values on the color bars indicate the number of samples in the bins and, if multiplied by 10, provide the probability density. Observing histograms (a) and (b) corresponding to subnet A, it is found that the spread of the results is greater in the case of DS3, which translates into a lower prediction efficiency than the estimation efficiency for DS1. In addition, for WF values close to the maximum, the predictions for DS3 considerably underestimate the true values. Regarding subnets B (c,d) and C (e,f), similar behaviors were found: slightly better for DS1 in the case of B and, surprisingly, slightly better for DS3 in the case of C. The histograms (g,h) of subnet D, magnified four times in the insets of the figures, show a higher concentration and lower dispersion of the samples from DS3 in the vicinity of the diagonal line. In addition, for WF values close to zero, subnet D has a systematic tendency to overestimate.

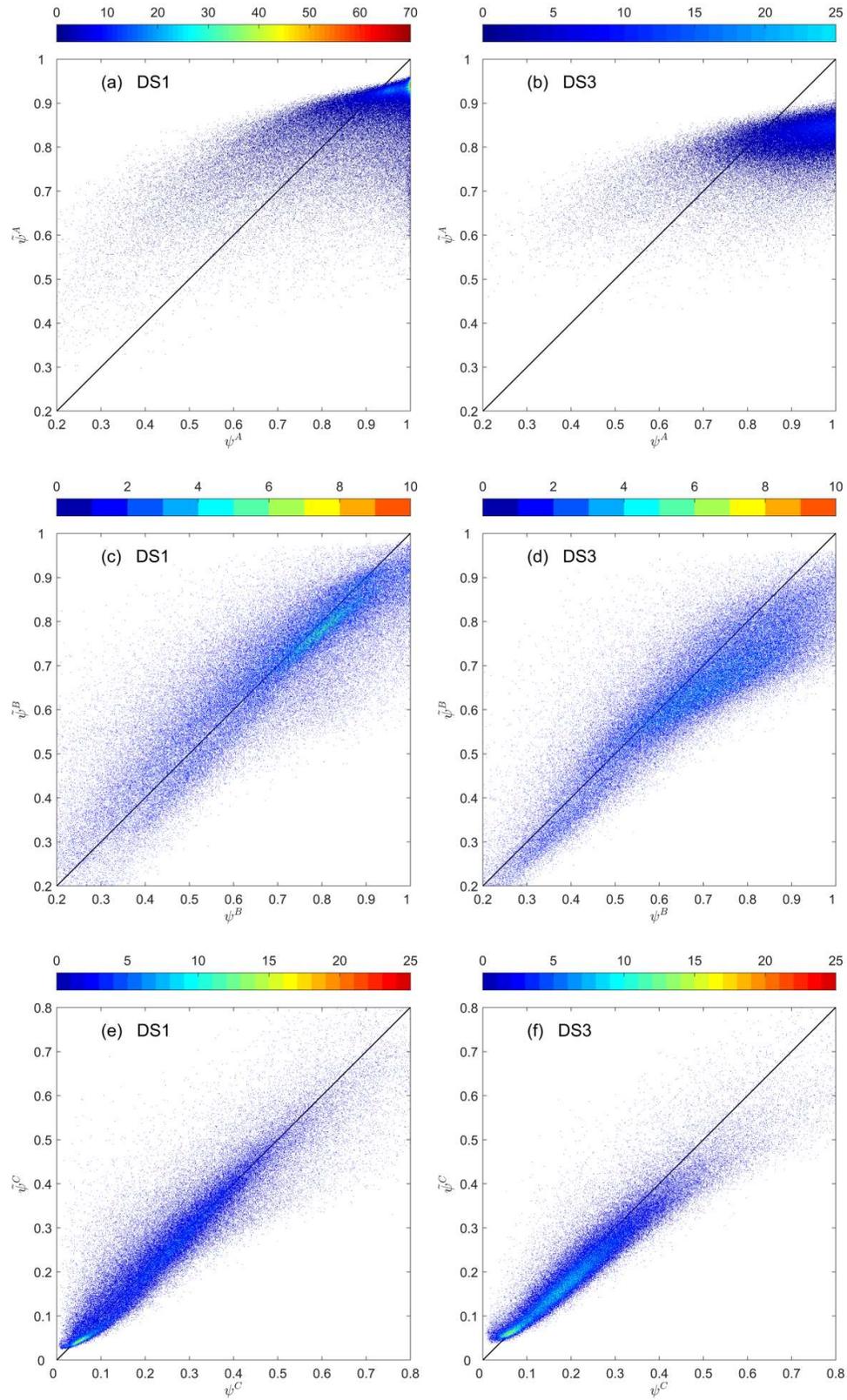

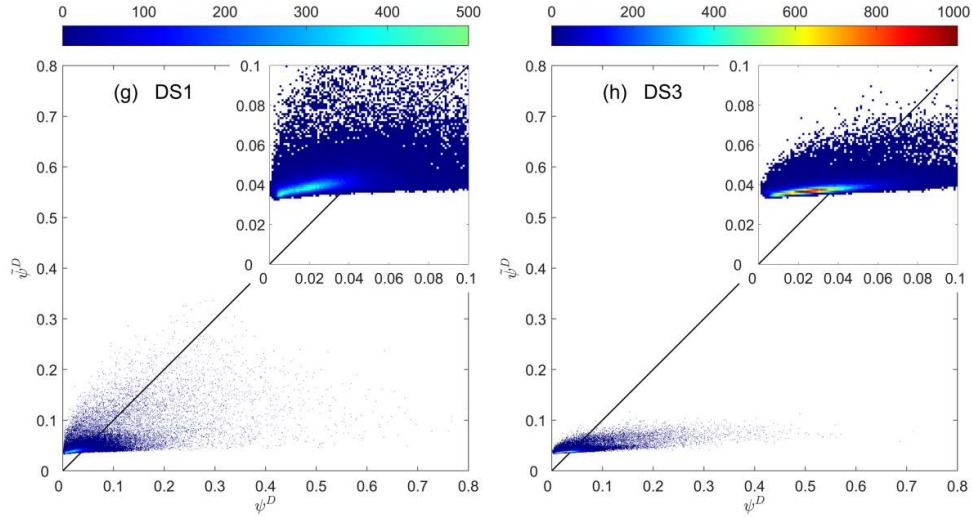

**Figure 8.** Estimated/predicted wave function values versus true solution values (given by the finite element method) for: node A (a,b), node B (c,d), node C (e,f), and node D (g,h). The diagonal lines represent the ideal, perfect prediction. The color scale is a measure of the in-plane density of samples. Histograms (a,c,e,g) present the estimation made on the training DS1, while (b,d,f,h) show the prediction for the testing DS3.

As explained above for the learning graphs, the predictive efficiency of the subnets for which the typical values of the WF approach the ends of the interval (0,1) is limited by the decrease in the logistic sigmoid derivative of the output neurons.

It was mentioned in Section 2.3 that the subnets are not correlated during training. An attempt to improve the method would require the encoding of the SE itself in the mathematical model of the network, that is, obtaining a physics-informed NN [50]. Because the partial derivatives in the equation assume variations with the coordinates of the WFs, it is clear that such an approach must correlate the training of the subnets related to the energy and neighboring nodes. Presumptively, the advantages would be more accurate predictions of the WFs and a better extrapolation of the method for inputs that are significantly different from those in the training DSs. However, these improvements will come with the cost of more calculations per iteration and a longer training time.

### 3.5. Training and testing the neural network. Calculation of quantitative indicators

After training on DS1 the entire NN formed by $M = 2764$ subnets for estimating the WF and one subnet for estimating the energy, the accuracy indicators can be calculated and compared. For this, the trained network was provided with all input data (sample CPs) from the three DSs, and the estimated/predicted results were statistically analyzed.

Figure 9 shows the results obtained in the form of superposed histograms. The training DS1 histogram was plotted as a contour line to demonstrate how well it matched the validation DS2 histogram for each indicator. It should be noted that although individually different, as shown in the Appendix, the CPs in DS2 are related to those in DS1 by the nature of their generating algorithm. However, the functions of DS3 are much more dissimilar to those of DS1 and DS2.

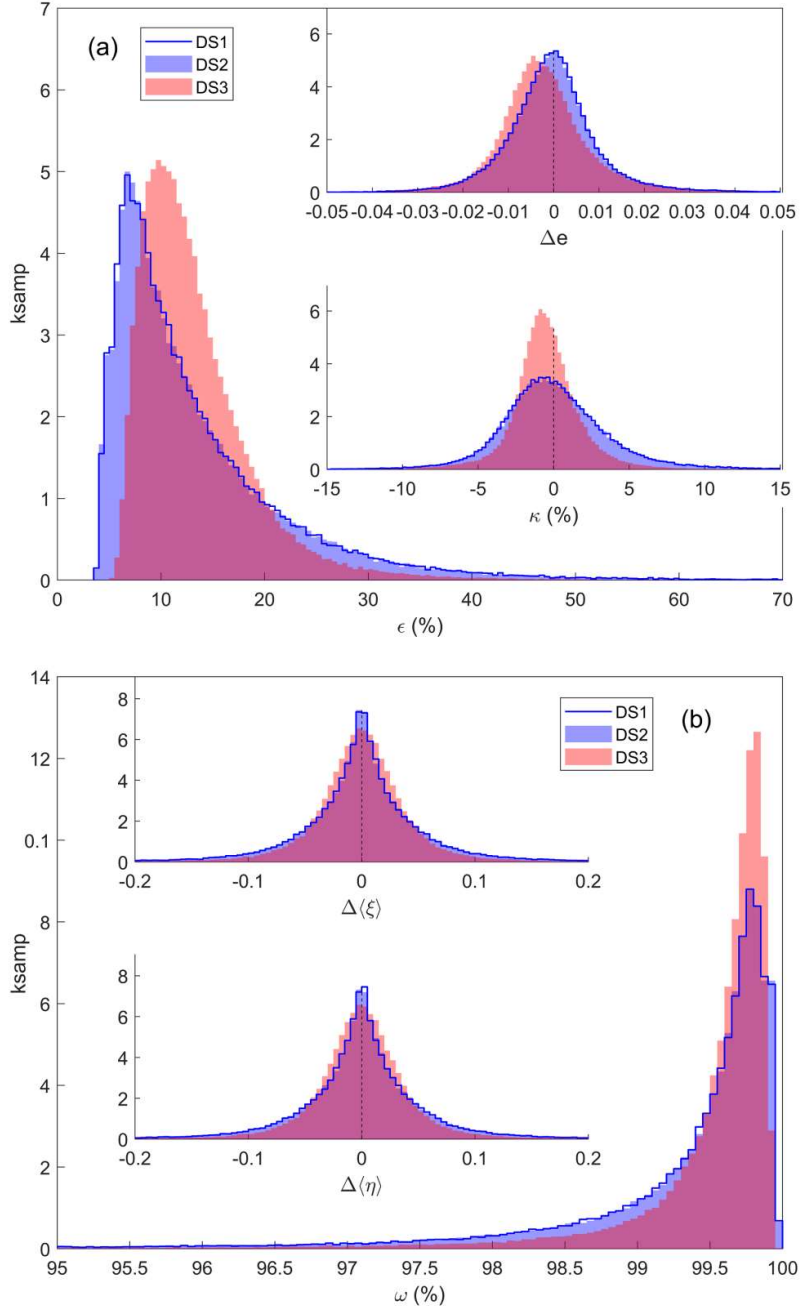

**Figure 9.** Multiple histograms comparing the neural network estimations/predictions for all 300 ksamp in the training DS1, validation DS2 and testing DS3: (a) relative difference between the estimated wave function $\tilde{\psi}_\sigma$ and the true solution $\psi_\sigma$; upper inset: deviation of the estimated/predicted energy from the true value; lower inset: energy relative deviation; (b) relative spatial overlap of estimated/predicted wave functions; insets: deviations of average $\xi$ and $\eta$ positions.

Comparing the results presented in Fig. 9 shows that even though the accuracy is slightly lower for DS3, the NN still behaves very well for a different CP type. For DS2, the estimation is better than for DS3 because nothing based on some classical algorithm is absolutely "arbitrary" and therefore the NN not only "learns" the Schrödinger problem, but also "learns" particular patterns encountered in the training DS1. From Fig. 9a, it can be seen that the relative difference between the predicted and true WFs is somewhat larger for DS3, which shifts the histogram profile to the right as compared to the DS1 histogram. Thus, the maximum of the DS3 histogram is shifted to the right by approximately 3%, but it is approximately the same height. From the insets of Fig. 9a, it appears that for DS3, the energy subnet has a greater tendency to underestimate. This is intuitive because the average value of the energies in DS3 was considerably higher than those in DS1 and DS2, as shown in Fig. 3. Figure 9b demonstrates that the true and estimated WFs overlap spatially in a proportion of over 98% for the vast majority of samples, and to an even greater extent for DS3. This can be explained by the stronger nature of the specific confinement of DS3, as demonstrated by the higher energies in this set. Stronger confinement means that functions are constrained in a narrower spatial region, resulting in better overlap. The insets of Fig. 9b show histograms of the average position differences and are almost identical, which was expected given that the algorithms generating the randomized CPs do not favor any particular direction.

### 3.6. Neural network predictions for several symmetric confinement potentials

The NN accuracy indicators in particular cases of symmetrical CPs (which cannot be found in any of the three DSs) were calculated. These cases were also used to plot and compare the aspect of the WFs predicted by the NN with the expected WFs calculated by the FEM. The following analytically defined 2-D finite-barrier potential wells were considered, all with equal confining zone areas:

a) $v(\xi, \eta) = H(\sqrt{\xi^2 + \eta^2} - 1)$, with $H$ denoting the Heaviside step function;

b) $v(\xi, \eta) = v_{min} + (1 - v_{min})H(\sqrt{\xi^2 + \eta^2} - 1)$, where $v_{min} = 1/4$;

c) $v(\xi, \eta) = H\left(\sqrt{\frac{\xi^2}{a^2} + \frac{\eta^2}{b^2}} - 1\right)$, where $ab = 1$ and eccentricity $e = \sqrt{1 - \frac{b^2}{a^2}} = \frac{7}{10}$;

d) $v(\xi, \eta) = \min(\xi^2 + \eta^2, 1)$;

e) $v(\xi, \eta) = H\left(|\xi| + |\eta| - \sqrt{\frac{\pi}{2}}\right)$.

It should be noted that CP (b) does not have the same scale factor as the others in the sense defined in the theory section; however, even in this case, the NN makes a good prediction. Figure 10 presents the CPs, solutions given by the FEM, and corresponding predictions of the NN. All surface plots were prepared via 2-D interpolation of the scattered function values at the nodes of the mesh subset used to train the network. It is observed that the NN solutions are not as smooth as the exact solutions, and generally have slightly smaller amplitudes.

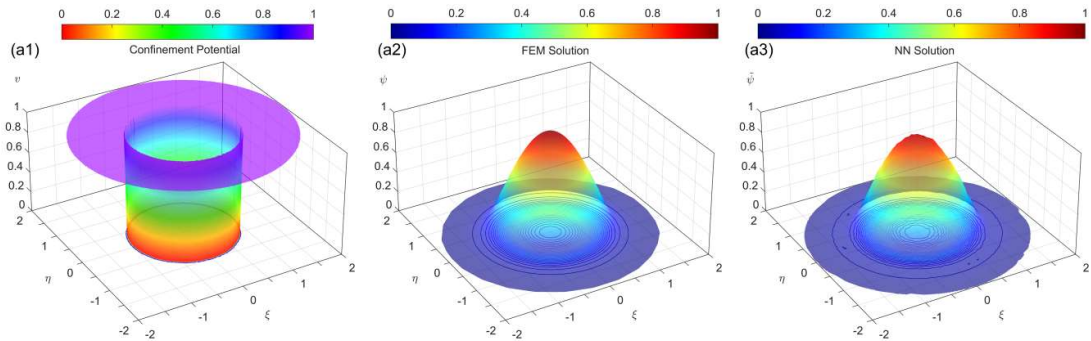

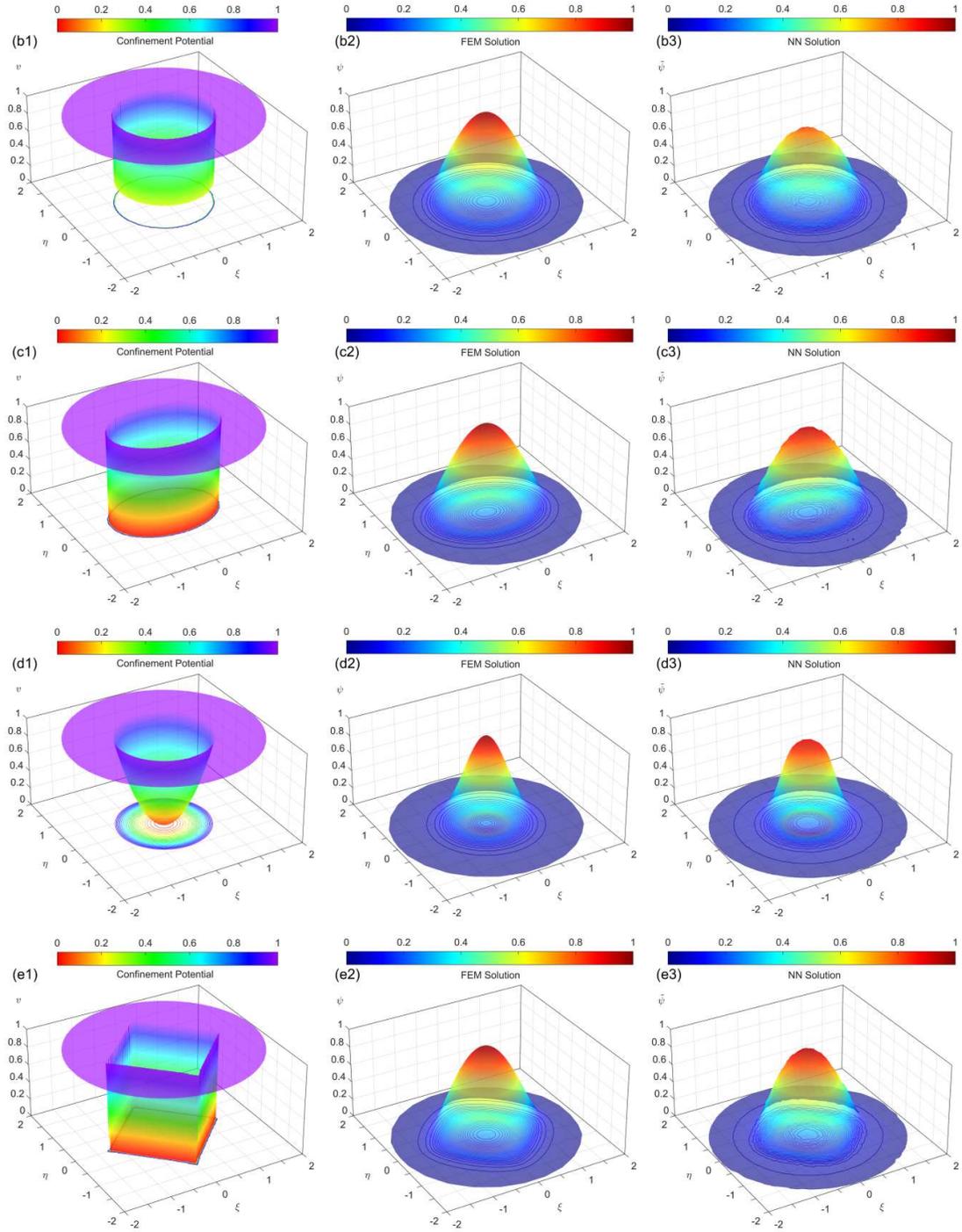

**Figure 10.** Confinement potentials, finite element method solutions and neural network predictions in five particular cases with analytically-defined, symmetrical potential functions. The semi-transparent surface plots allow the contour plots of isolines to be observed in the plane $(\xi, \eta)$. Quantum wells (a) and (b) are cylindrical confinements of depths 1 and 3/4, respectively; (c) has an elliptical confinement perimeter; (d) is a confinement with variable parabolic depth, and (e) has a square perimeter of confinement.

A quantitative assessment of the prediction accuracy of the NN in these cases is presented in Table II. Columns 3-10 of the table contain, from left to right: the WF relative deviation, WF relative spatial overlap, deviation of the average $\xi$ position, deviation of the average $\eta$ position, exact value of the energy, predicted value of the energy, deviation of the energy, and relative deviation of energy. The best value in each indicator category is underlined, and the worst value is italicized.

Table II. Accuracy indicators for the quantum wells presented in Fig. 10

| Case | Potential | $\epsilon$(%) | $\omega$(%) | $\Delta\langle\xi\rangle$(E-4) | $\Delta\langle\eta\rangle$(E-4) | $e$ | $\tilde{e}$ | $\Delta e$ | $\kappa$(%) |
|---|---|---|---|---|---|---|---|---|---|
| (a) | | <u>3.72</u> | <u>99.95</u> | -4.45 | <u>-7.29</u> | 0.1541 | 0.1560 | 0.0019 | 1.23 |
| (b) | | *14.58* | 99.93 | -2.71 | -10.41 | 0.3959 | 0.3904 | *-0.0055* | -1.39 |
| (c) | | 5.80 | 99.90 | *5.68* | *-22.37* | 0.1589 | 0.1598 | <u>0.0009</u> | <u>0.57</u> |
| (d) | | 11.92 | *99.63* | 2.41 | 15.80 | 0.3906 | 0.3929 | 0.0023 | 0.59 |
| (e) | | 4.91 | 99.92 | <u>-0.92</u> | -18.68 | 0.1609 | 0.1637 | 0.0028 | *1.74* |

There were no double QW samples in any of the DSs. However, systems where the potential evolves from a single well to a double-well structure are very interesting for the study of quantum phase transitions [51,52] and the design of quantum gates [53]. Recent studies have investigated whether ML methods can be applied to predict the onset of a quantum phase transition and extrapolate the properties of the system in the phase separated from the training data by the phase transition [25]. It is interesting to investigate whether the NN from the present study can make reasonable predictions on a double-well configuration, given that it has not "met" any double-well during training. For this, we divided the cylindrical confinement (a) from Table II into two equal semi-cylindrical wells by a potential barrier of variable width $w_b$. The width was gradually increased from 0 (cylindrical confinement) to 3/4 of the confinement radius (in which case the two wells can be understood as almost distinct systems). The identical wells were distanced without changing the initial total confinement area. We calculated the ground energy of the double QW using both methods, as shown in Fig. 11.

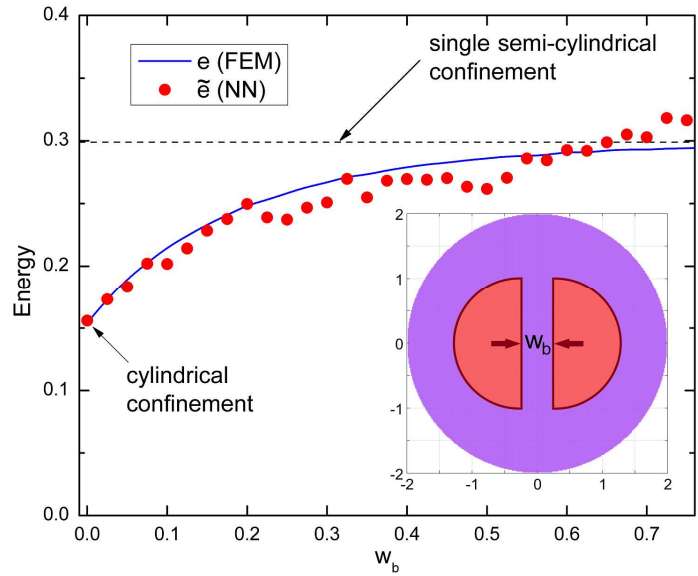

Figure 11. Transition from single cylindrical to double semi-cylindrical quantum structure with the same total confining area. Dashed line shows the ground energy level of a single semi-cylindrical quantum well.

The solid line represents the FEM solution, with the typical asymptotical increase from the energy of the cylindrical well to almost twice that value for the single semi-cylindrical well of the halved area. The circular symbols represent the NN predictions that match the exact solution well, except for the intrinsic numerical fluctuations in the method. For $w_b < 0.6$ there is a tendency for the NN to underestimate the energy, but for larger values of $w_b$ it is expected to overestimate.

Finally, we discuss the possibility of extending this neural model to the 3-D Schrödinger problem, provided that the parallel computation resources required for training are available. Concerning the NN architecture, no qualitative change is necessary because there is nothing intrinsically "two-dimensional" about the IL or OL of the network. The input/output nodes are values of the confining potential/wave function in an indexed set of points, which can be distributed on a line (1-D mesh), surface (2-D mesh), or volume (3-D mesh). Quantitatively, however, there are differences in terms of the number of nodes in the IL and OL and implicitly the computational effort required to train the network. The number of nodes involved must be adequate for the complexity of the 3-D confinement functions to ensure a satisfactory predictive performance. This aspect is not surprising, because even higher-dimensional FEMs involve meshes with a large number of nodes and high computation times. Therefore, a natural extension of our method is to address 3-D problems specific to quantum dots.

## 4. Conclusions

In this study we used a deep learning technique to approach SE in 2-D QWs with finite walls and random CPs. An NN with two HLs and 2764(+1) subnets was trained using a set of CPs and their corresponding WFs and energies, which were previously calculated using a FEM. An important advantage of this NN architecture is that it is easy to parallelize calculations. Several accuracy indicators have been proposed to test NN predictions. The network was trained on the DS1 containing 100 ksamp with the SGD method and the training was validated with respect to the equally sized DS2. The network was also applied to a different testing DS3 of 100 ksamp and the results were compared by using the accuracy indicators. It was found that the network has good prediction accuracy, which is slightly lower for the test set than for the training and validation sets, as expected. Several cases with analytical CP have also been approached, presenting explicit graphs of potentials and predicted WFs and listing the accuracy indicators. The ability of the NN to make predictions on the ground state energy of a double quantum well has been demonstrated.

The improvements and developments that can be made in future studies are as follows: (i) using adaptive learning parameters, (ii) generalization of the NN solutions to 3-D problems (quantum dots), (iii) predicting energies and WFs of excited levels, and (iv) improving the method in the context of physics-informed NNs.

**Funding Statement:** NO funding to declare.
**Conflict of Interest Declaration:** The authors declare that they have NO affiliations with or involvement in any organization or entity with any financial interest in the subject matter or materials discussed in this manuscript.
**Author Contributions:** AR prepared the datasets and implemented the neural network. AR and CAD contributed to the training, validation, and testing of the neural network. AR and CAD contributed to the analysis of the results and to the writing of the manuscript.

**Appendix**

Confining potential samples from DS1/DS2 and DS3 were generated using random-number-based algorithms. There are an infinite number of ways to accomplish this, and any chosen method has a large dose of arbitrariness. Setting up the finite-wall confining potential involves the completion of two main stages: i) the random definition of a confining perimeter outside which the function $v$ has a constant value 1, and ii) the random definition of the potential function inside the confining perimeter. The first stage is common to the three DSs, unlike the second stage, where different methods were used for DS1/DS2 and DS3. In the following, we briefly explain the two stages of the algorithms.

i) For each potential sample, an integer $n$ was randomly chosen such that $3 \leq n \leq 7$, the possible values occurring with equal probabilities. $n$ represents the number of coplanar distinct points $\{K_p\}_{1 \leq p \leq n}$ used to further define the confinement perimeter. The circumradius $r_{c,n} = \sqrt{2\pi/n \csc(2\pi/n)}$ of the regular polygon with $n$ vertices and fixed area $\pi$ was used to randomly define the radial position $\rho_p$ of each vertex $K_p$ in the interval $(0.9r_{c,n}, 1.1r_{c,n})$, with a uniform probability distribution. If $n > 4$, the radial position of a single randomly chosen vertex is halved with a probability of $1/2$, to allow slightly concave confining perimeters. The random angular positions of the vertices were defined with respect to the positive $\xi$ semi-axis using the formula $\theta_p = (p-1)\, 2\pi/n + \delta\theta_p + \theta_0$. Here, $\delta\theta_p$ is randomly and uniformly distributed between 0 and $\pi/n$ for each vertex in part, and $\theta_0$ is randomly and uniformly distributed between 0 and $2\pi$, and is the same for all the vertices. The confining perimeter was obtained by the closed interpolation of the set $\{K_p\}$, with Cartesian coordinates $(\rho_p \cos\theta_p, \rho_p \sin\theta_p)$. With a probability of $1/3$, interpolation was performed using one of the following three methods: linear, spline, or piecewise cubic Hermite polynomial.

ii) For DS1/DS2, the algorithm was based, with a probability of $5/8$, on various combinations of up to eight random harmonic terms depending on the coordinates $\xi$ and $\eta$. With a probability of $3/8$, the algorithm uses combinations of piecewise linear interpolation functions or piecewise cubic Hermite interpolation polynomials. Interpolations were performed with a random number of three to nine randomly chosen sample points with a uniform probability distribution inside the confining zone. The functions generated in this manner were additionally modulated in amplitude with random fractional power-type envelope functions or constant functions on intervals. For DS3, the algorithm is based on a linear interpolation of up to 100 sample points scattered randomly and uniformly inside the confinement domain. For all the DSs, the query points of the interpolations were the nodes of the $\Xi_N$ mesh lying inside the confinement perimeter.

The 2-D color gradient plots in Figs. A1 and A2 show a number of 20 random examples of confinement functions $v(\xi, \eta)$ generated with the algorithm specific to sets DS1/DS2 and DS3, respectively.

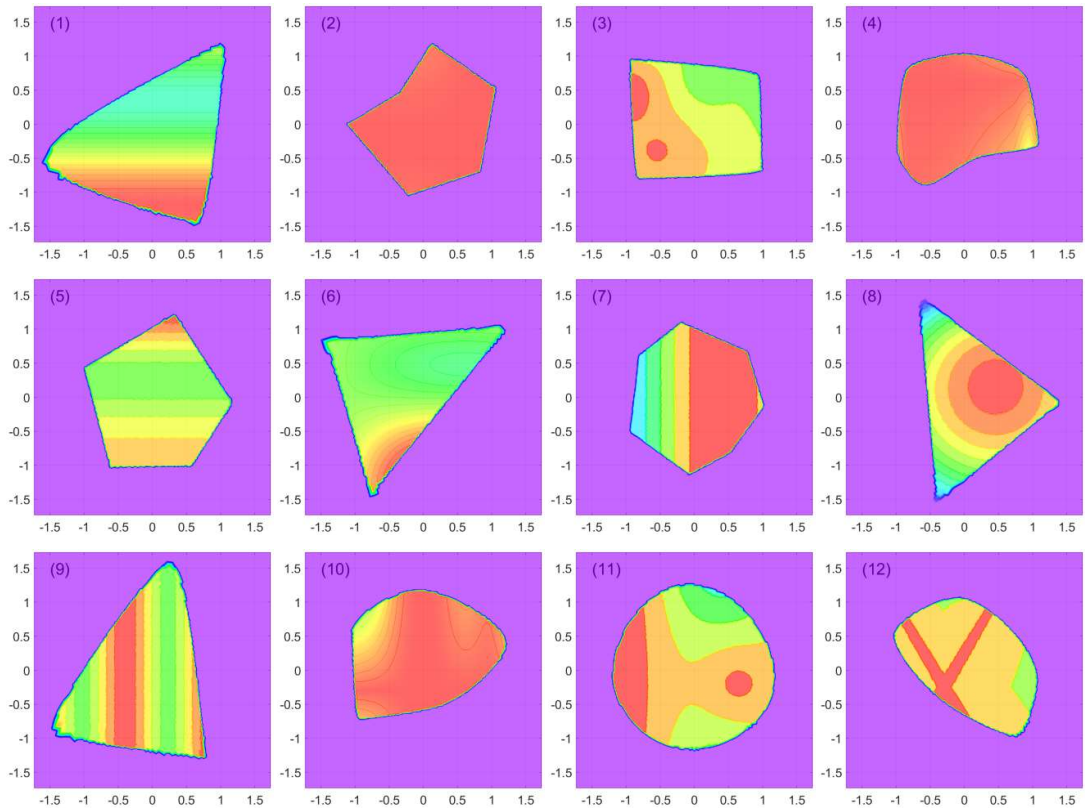

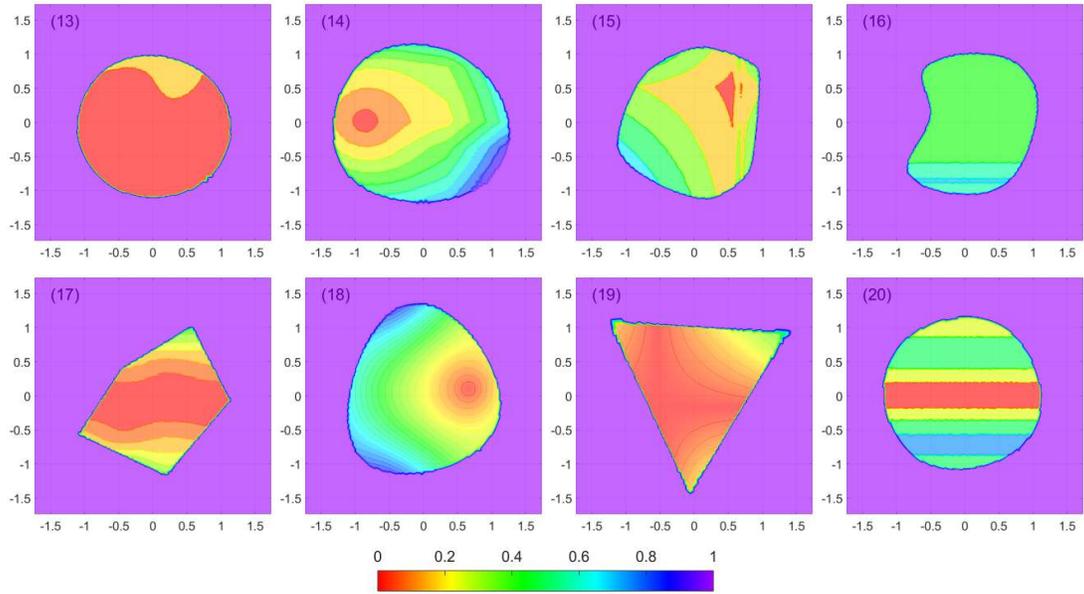

**Figure A1.** Several random DS1/DS2-type confining potentials $v_\sigma(\xi,\eta)$. Samples $\sigma = \overline{1,10}$ are from the training DS1 and samples $\sigma = \overline{11,20}$ are from the validating DS2.

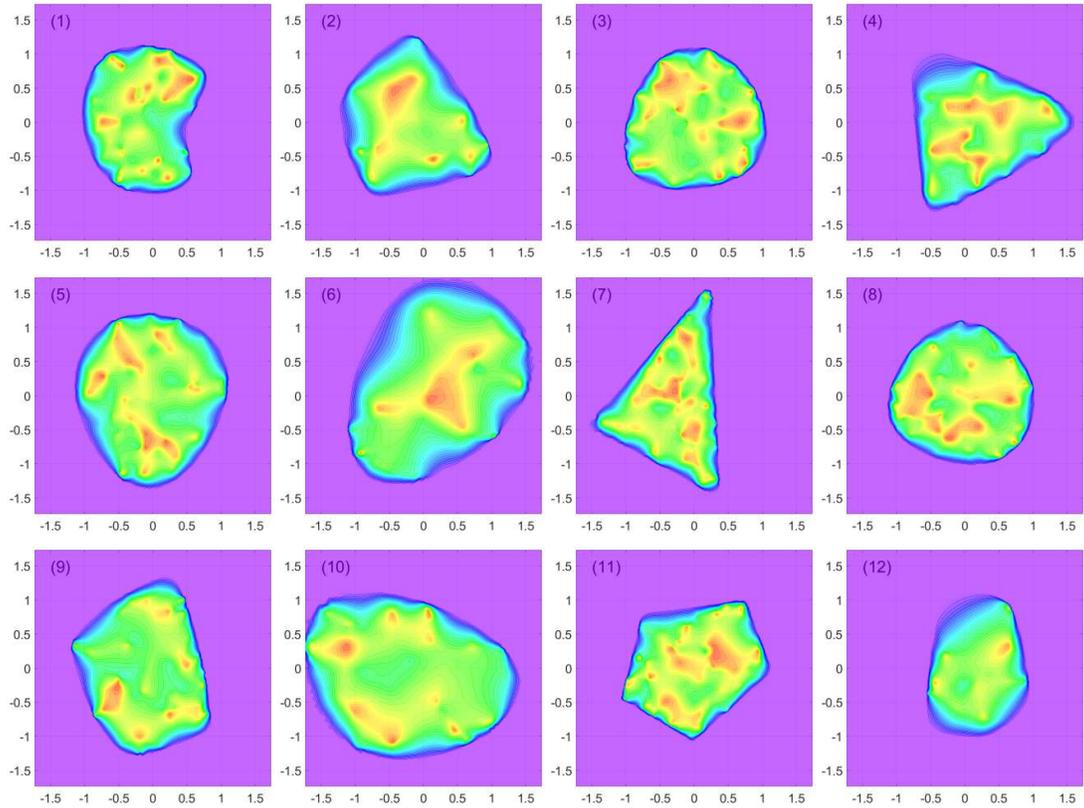

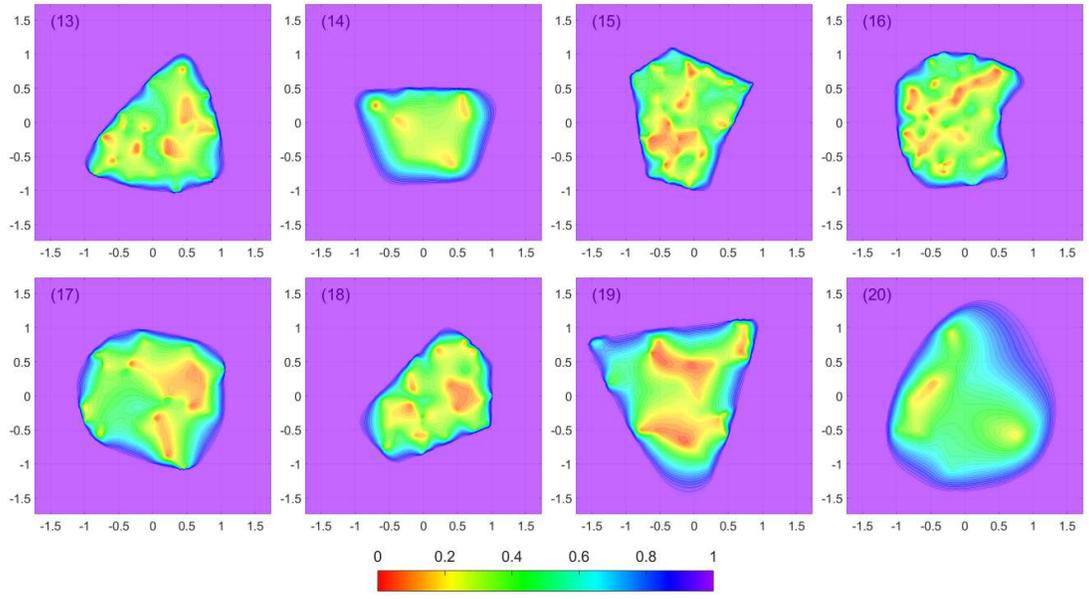

**Figure A2.** Several random testing DS3 confining potentials $v_\sigma(\xi, \eta)$.